\documentstyle[prd,aps,eqsecnum]{revtex}

\begin{document}

\title{Age of the Universe: Influence of the Inhomogeneities on the
global Expansion-Factor}

\author{
Heinz Russ$^{1}$, Michael H. Soffel$^{2}$,  Masumi Kasai$^{3,4}$
and Gerhard B\"orner$^{4}$}

\address{\small\it
${^1}$ Institut f\"{u}r Astronomie und Astrophysik,
Universit\"{a}t T\"{u}bingen, Auf der Morgenstelle 10, 
72076 T\"{u}bingen, Germany
  \\
${^2}$ Institut f\"ur Planetare Geod\"asie, Universit\"at Dresden,
Mommsenstr. 13, 01082 Dresden, Germany
   \\
${^3}$ Department of Physics, Hirosaki University, 3 Bunkyo-cho, 
Hirosaki, 036 Japan
  \\
${^4}$ Max-Planck Institut f\"ur Astrophysik, Karl-Schwarzschild
Str. 1, 85740 Garching, Germany
}

\date{26. May 1997}

\maketitle


\begin{abstract}
For the first time we calculate
quantitatively the influence of 
inhomogeneities on the global expansion factor by 
averaging the Friedmann equation. In the framework of the
relativistic second-order Zel'dovich-approximation scheme
for irrotational dust we use observational results in form of
the normalisation constant fixed by the COBE results and we check
different power spectra, 
namely for adiabatic CDM, isocurvature CDM, HDM, WDM,
Strings and Textures. 
We find that the influence of the inhomogeneities on the
global expansion factor is very small.
So the error in determining the age of the universe using the
Hubble constant in the usual way is negligible.
This does not imply that the effect is negligible for local
astronomical measurements of the Hubble constant.
Locally the determination of the redshift-distance relation
can be strongly influenced by the peculiar velocity fields
due to inhomogeneities. Our calculation does not consider
such effects, but is contrained to comparing globally homogeneous
and averaged inhomogeneous matter distributions.
In addition we relate our work to previous treatments.
 
\end{abstract} 

\pacs{}

\section{Introduction} \label{sec:intro}

Lower limits of the age of the universe are observationaly determined in
many ways.
Measurements of isotopic ratios of radioactive nuclei
determine the ages of meteorites by $4.5 $ Gyrs 
\cite{bi:boerner,bi:kirsten}.
Studies of the cooling of white dwarfs \cite{bi:nather} lead to an age of
our solar system of about 1 Gyr.
Galactic ages could be determined to lie in a range of 
$12.6$ to $19.6$ Gyrs by measuring the abundance ratio of 
different isotopes of elements \cite{bi:boerner,bi:burbidge}.
Measurements of the luminosity of stars located at the turn-off point of
the Hertzsprung Rusell diagram determine the ages of globular clusters
to lie in a range of $12$ to $18$ Gyrs 
\cite{bi:ibnren,bi:chappi,bi:bolte}.
On the other hand upper limits of the age of the universe can be derived 
using cosmological models. Using a standard 
Friedmann-Lema\^{\i}tre-Robertson-Walker (FLRW) model with vanishing
cosmological constant ($\Lambda =0$), the inverse of the Hubble
parameter as measured today, $H_0^{-1}$,
provides an upper limit of the age of the universe
(the index 0 indicates values at the present time).
Recent measurements of cepheid variables in the virgo cluster
\cite{bi:freed,bi:pierce} lead to a Hubble parameter of about 
$H_0=80$ km/(s$\cdot$Mpc)
(upper limit $H_0^{-1}\approx 12.2$ Gyrs),
leading to an age of $8.15$ Gyrs
for a flat universe with vanishing cosmological constant.
 This is far below the
observational lower limits cited above. There are several ways out of this
dilemma:

The first is believing in a lower value of the Hubble parameter.
There are two reasons for that: firstly there exist other
observational results \cite{bi:saha}, secondly the redshift-distance
relation can be influenced by the inhomogeneities, which will
influence the Hubble constant.

The second is to believe that the high value of the Hubble constant
comes from the fact that we live in an underdensed region of the universe,
whereas on average over the whole universe the expansion parameter is smaller
\cite{bi:shi}.

The third is believing in some nonvanishing cosmological constant 
\cite{bi:freed,bi:priester},
where under some circumstances no upper limit
can be derived ($t_0>H_0^{-1}$),
and the age of the universe can be about $30$ Gyrs or 
even higher \cite{bi:priester}.

In this paper we want to investigate still another way.
In the usual calculation of its
age, the universe is assumed to be exactly isotropic and homogeneous. This
might be a good approximation due to the high isotropy of the 
microwave-background-radiation, so that the FLRW description might be valid
in some averaged sense. On the other hand the inhomogeneities are large,
even on large scales, for example at a scale of $\sim 10$ Mpc
the density constrast $\delta=\delta\rho/\rho$ might reach unity.
In the early universe deviations from homogeneity and isotropy were
small, but after the deviations became nonlinear, these inhomogeneities could
influence the global expansion factor. This effect is called by us
the backreactions
of the inhomogeneities. As a result of these backreactions the value of the
Hubble parameter cannot be taken in the usual way for a determination
of the age of the universe. 
We will calculate quantitatively the effect of the backreactions
and we will see how large the deviations from the usual 
age-determinations are.
This paper is organized as follows:
In Sec.~II we present the basic equations and the averaged Friedmann
equation in a general form.
In Sec.~III we use the results of the relativistic Zel'dovich type
approxomation to second order (Russ {\it et al.} \cite{bi:russ})
 based on the tetrad formalism in cosmology
(Kasai \cite{bi:kasai95})
and calculate the backreactions using
different power spectra and the normalisation constant fixed by the
COBE results.
This paper was influenced by the pioneering paper from
Bildhauer and Futamase \cite{bi:bild}; we compare our results with theirs and
others (Buchert, Ehlers \cite{bi:buchehlers,bi:ehlersbuch}, 
Futamase \cite{bi:futa,bi:futarev}) in Sec.~IV.
Sec.~V is devoted to conclusions.

\section{The Friedmann equation in an inhomogeneous universe}

In this section, we summarize a general relativistic
treatment to describe the non-linear evolution of an 
inhomogeneous irrotational\footnote{We do not consider here the
effect of rotation, which might turn the effect of the inhomogeneities
in the opposite direction, i.e. tending to increase the age of the universe.}
universe
\cite{bi:kasai95,bi:kasai1,bi:kasai93}.
The models we consider contain irrotational dust with 
energy-density $\rho$ and four-velocity $u^{\mu}$. We
will neglect the curvature constant $k$ and a
possible cosmological constant $\Lambda$.
Neglecting the fluid pressure and the vorticity is a 
reasonable assumption in a cosmological context.
In comoving synchronous coordinates, the
line element can be written in the form
(Indices $\mu, \nu, \cdots$ run from $0$ to $3$ 
and indices $i, j, \cdots$ run from $1$ to $3$)
\begin{equation}
ds^2 = -c^2 dt^2 + g_{ij} dx^i dx^j
\end{equation}
and $u^{\mu}=(c,0,0,0)$.
Then, Einstein's field equations read
\begin{eqnarray}
\label{eq:G00}
\frac{1}{2} \left[\;{^3\!R^i_{\ i}}\;c^2 + \left( u^i_{\ ;i} \right)^2 
                          - u^i_{\ ;j} u^j_{\ ;i}\;\right]
  &=& \frac{8\pi G}{c^2}\rho \;,\\ 
\label{eq:G0i}
u^i_{\ ;j \|i} - u^i_{\ ;i \|j}&=&0 \;,\\ 
\label{eq:Rij}
\dot{u}^i_{\ ;j} + u^{k}_{\ ;k} u^i_{\ ;j} + {^3\!R}^i_{\ j}\;c^2 
  &=& \frac{4\pi G}{c^2}\rho\, \delta^i_{\ j} \;,
\end{eqnarray}
where ${^3\!R}^i_{\ j}$ is the three-dimensional 
Ricci-tensor,
\begin{equation}\label{excurv}
u^i_{\ ;j} = \frac{1}{2} g^{ik}\dot g_{jk}
\end{equation}
is the extrinsic curvature, $\|$ denotes the covariant derivative with
respect to the three metric $g_{ij}$, and an overdot ($\dot{}$)
denotes $\partial/\partial t$.
We introduce the conformal factor $a(t)$ by
\begin{equation}
g_{ij} = a(t)^2 \gamma_{ij}
\end{equation}
and we introduce the quantity $V^i_{\ j}$, describing
the deviation from a homogeneous and isotropic expansion
\begin{equation}
V^i_{\ j} \equiv u^i_{\ ;j} - 
\frac{\dot a(t)}{a(t)} \delta^i_{\ j}
=\frac{1}{2} \gamma^{ik} \dot\gamma_{jk}\;.
\end{equation}
Then we can write two of the Einstein equations in the following 
form, which we call the Friedmann equations
\begin{equation}\label{eq:dota}
\frac{{\dot a}(t)^2}{a(t)^2}=
\frac{8\pi G}{3c^2}\rho 
-\frac{c^2}{6}\,{^3\!R}
-\frac{1}{6}\Bigl((V^k_{\ k})^2-V^l_{\ k}V^k_{\ l}\Bigr)
-\frac{2\dot a(t)}{3a(t)}V^k_{\ k}
\end{equation}
and
\begin{equation}\label{eq:ddota}
\frac{\ddot a(t)}{a(t)}=
-\frac{4\pi G}{3c^2}\rho 
-\frac{1}{3}V^l_{\ k}V^k_{\ l}
-\frac{1}{3}\dot V^k_{\ k}
-\frac{2\dot a(t)}{3a(t)}V^k_{\ k}\;.
\end{equation}
We introduce the following averaging procedure \cite{bi:explain}
\begin{equation}
<A>=\frac{1}{V}\int_{V} A \sqrt{g}\, d^3x\;,
\end{equation}
where $g\equiv {\rm det}g_{ij}$.
$V$ is the comoving volume of a compact domain ${\cal D}(t)$ of the
fluid \cite{bi:buchehlers,bi:ehlersbuch}.
$V$ should
be sufficiently large so that we can assume periodic boundary 
conditions. The scale factor $a_D(t)$ describes the
expansion of this volume. Therefore the expansion rate of
the universe is defined by
\begin{equation}
3\frac{\dot a_D(t)}{a_D(t)}\equiv\frac{\dot V}{V}
=3\frac{\dot a(t)}{a(t)}+<V^k_{\ k}>\;.
\end{equation}
We then average the Friedmann equations, apply the
commutation rule and neglect higher order terms
(see APPENDIX \ref{A}) to get the averaged Friedmann equations in the form
\begin{equation}\label{eq:dotaD}
\frac{\dot a_D(t)^2}{a_D(t)^2}=
\frac{8\pi G}{3c^2}<\rho>
-\frac{c^2}{6}<{^3\!R}>
-\frac{1}{6}<(V^k_{\ k})^2-V^l_{\ k}V^k_{\ l}>
\end{equation}
and
\begin{equation}\label{eq:ddotaD}
\frac{\ddot a_D(t)}{a_D(t)}=
-\frac{4\pi G}{3c^2}<\rho>
+\frac{1}{3}<(V^k_{\ k})^2-V^l_{\ k}V^k_{\ l}>\;.
\end{equation}
These are the general equations for the evolution of the
expansion factor of an inhomogeneous universe. They do not
depend on a specific model of the universe. 
Eq.~(\ref{eq:ddotaD}) has already been discovered by
Buchert and Ehlers \cite{bi:buchehlers,bi:ehlersbuch}
(see Sec.~IV).
\section{Model of the inhomogenous universe}
We use the solution of the
relativistic Zel'dovich approximation to second order
(Russ {\it et al.} \cite{bi:russ}) based on the tetrad formalism
(Kasai \cite{bi:kasai95}) to get
\begin{equation}\label{solu}
\frac{\dot a_D(t)^2}{a_D(t)^2}=
\frac{8\pi G}{3c^2}\rho_b(t)
-\frac{1}{t^2_{\rm in}}\int_{V}\frac{100}{243\,a_D^2\,c^2 t^2_{\rm in}}
\Psi^{,k}\Psi_{,k} d^3x\;.
\end{equation}
The function $\Psi({\bf x})$
is related to the initial displacements of the particles,
to first order it represents the potential of the density fluctuations
$-\Psi^{,k}_{\ ,k}=\delta({\bf x}, t_{\rm in})$. Here we put 
$a_D(t_{\rm in})=1$ and $V(t_{\rm in})=1$.
For a justification of eq.~(\ref{solu}) see APPENDIX \ref{B}.
We use the background relationships
\begin{equation}
\rho_b(t_{\rm in})=\frac{c^2}{6\pi G t^2_{\rm in}}
\qquad {\rm and} \qquad
t_{\rm in}=\frac{2}{3H_0(1+z_{\rm in})^{3/2}}\;,
\end{equation}
where $H_0$ is the present value for
the Hubble parameter
\begin{equation}
H_0^2\equiv\frac{\dot a_D^2(t_0)}{a_D^2(t_0)}=
\frac{8\pi G}{3c^2}\Bigl(\rho_b(t_0)+\rho_{\rm corr}(t_0)\Bigr)
\equiv \frac{8\pi G}{3c^2}\rho_b(t_0)\Bigl(1+\delta_{\rm corr}(t_0)\Bigr)\;.
\end{equation}
We then get
\begin{equation}\label{eq:friedcomp}
\frac{\dot a_D^2(t)}{a_D^2(t)}=\frac{H_0^2}{1+\delta_{\rm corr}(t_0)}\left(
\frac{a_D^3(t_0)}{a_D^3(t)}+\frac{a_D^2(t_0)}{a_D^2(t)}\delta_{\rm corr}(t_0)
\right)
\end{equation}
with
\begin{equation}
\delta_{\rm corr}(t_0)=-\frac{25}{108}
10^{-6}h_0^2(1+z_{\rm in})^4{\rm Mpc}^{-2}
\int_{V}\Psi^{,m}\Psi_{,m}d^3x
\end{equation}
and $H_0=100 h_0$ km/(s$\cdot$Mpc).
We integrate the Friedmann equation (\ref{eq:friedcomp})
to get the age of the universe:
\begin{equation}
t_0=\frac{2}{3}\frac{\sqrt{1+\delta_{\rm corr}(t_0)}}{H_0}\left(
\frac{3}{2}\int_{0}^{1}
\frac{\sqrt{x}dx}
{\sqrt{1+x\delta_{\rm corr}(t_0)}}\right)\;.
\end{equation}
Since $\delta_{\rm corr}(t_0)$
 has a negative sign the age of the inhomogeneous universe is
less than the age of a corresponding homogeneous one calculated
with a given Hubble constant.
Here, we want to estimate these differences quantitatively.
The flat background allows a Fourier decomposition 
$\delta({\bf x},t_{\rm in})=\sum_{\bf k}\delta_{\bf k}e^{i{\bf kx}}$,
so we get
\begin{equation}
\int_{V}\Psi^{,m}\Psi_{,m}d^3x
=\sum_{\bf k}\frac{1}{{\bf k}^2}|\delta_{\bf k}|^2\,,
\end{equation}
where $|\delta_{\bf k}|^2$ is the power spectrum of density fluctuations
\begin{equation}
|\delta_{\bf k}|^2= P(k)\;.
\end{equation}
What we need is to know
the power spectrum at initial time $t_{\rm in}$,
where the fluctuations are still
linear and just start to move into the nonlinear regime.
We choose $z_{\rm in}=8$ \cite{bi:jing}.
The power spectrum evolves according to \cite{bi:bomo}
\begin{equation}
P(k,t_{\rm in})=\frac{1}{(1+z_{\rm in})^2}P(k,t_{\rm pr})T^2(k)\;,
\end{equation}
where $T(k)$ is a transfer function and
the primordial power spectrum is assumed to be
\begin{equation}
P(k,t_{\rm pr})=Ak^n\;.
\end{equation}
The normalisation constant is fixed by the COBE results
\begin{equation}
A_{\rm COBE}=\left(\frac{96\pi^2}{5}\right)\Omega_0^{-1.54}c^4H_0^{-4}
\left(\frac{Q_{\rm rms}}{T_{\gamma_0}}\right)^2\;.
\end{equation}
The COBE rms-fluctuation is given to be $Q_{\rm rms}=9.3\mu$K
\cite{bi:stark}, if we
take $n=1$ (scale invariant primordial power spectrum), 
which is the most reasonable value \cite{bi:scara}. The temperature
of the Microwave Background Radiation is $T_{\gamma_o}=2.73$K.
The volume will be taken large enough so that we
can convert the sum into an integral
\begin{equation}
\sum_{\bf k}\rightarrow\frac{1}{(2\pi)^3}\int d^3k
\end{equation}
to get
\begin{equation}
\delta_{\rm corr}(t_0)
=-\frac{25}{216\pi^2} 10^{-6}h_0^2(1+z_{\rm in})^2 A \int k T^2(k)dk\;.
\end{equation}
In the following we will assume $h_0=0.8$ and $\Omega_{\rm tot}=1$.
For the normalisation constant one finds
$A\approx 4.35\cdot 10^5\, {\rm Mpc}^4$. In the following we will calculate
the age of the universe using different transfer functions $T(k)$
and power spectra $P(k)$, where $AkT(k)^2=P(k,t_0)$.
\subsection{Adiabatic Cold Dark Matter Fluctuations}

The transfer function for adiabatic CDM fluctuations is given by 
\cite{bi:bardeen}
\begin{equation}
T_{\rm CDM, ad}(k)=\frac{{\rm ln}(1+2.34q)}{2.34q}
\left[1+3.89q+(16.1q)^2+(5.46q)^3+(6.71q)^4\right]^{-1/4}\;,
\end{equation}
where
\begin{equation}
q\equiv\frac{k\theta^{1/2}}{\Omega_{\rm CDM} h_0^2{\rm Mpc}^{-1}}\;.
\end{equation}
Here $\theta=\rho_{er}/(1.68\rho_{\gamma})$ is a measure of the ratio of the
energy density in relativistic particles (photons plus neutrinos) to
that contained in photons. 
We will set $\theta=1$, corresponding to three flavors of 
relativistic neutrinos plus the photons, and we will take $\Omega_{\rm CDM}=1$.
Varying $\Omega_{\rm CDM}$ will change the result only slightly.\\
The result is $\delta_{\rm corr}(t_0)=-2.50 \cdot 10^{-3}$ and
the age of the universe becomes
\begin{equation}
t_0 \approx 0.9995 \cdot \frac{2}{3H_0}\;.
\end{equation}
\subsection{Isocurvature Cold Dark Matter Fluctuations}
The transfer function for isocurvature CDM fluctuations reads 
\cite{bi:bardeen}:
\begin{equation}
T_{\rm CDM, isoc}(k)=\left[1+\frac{(40q)^2}{1+215q+(16q)^2(1+0.5q)^{-1}}
+(5.6q)^{8/5}\right]^{-5/4}\;,
\end{equation}
where
\begin{equation}
q\equiv \frac{k}{\Omega_{\rm CDM}h_0^2{\rm Mpc}^{-1}}\;.
\end{equation}
We again will take $\Omega_{\rm CDM}=1$.\\
The results is $\delta_{\rm corr}(t_0)=-5.63 \cdot 10^{-4}$
and the age of the universe reads
\begin{equation}
t_0 \approx 0.99989 \cdot \frac{2}{3H_0}\;.
\end{equation}
\subsection{Hot Dark Matter + Adiabatic CDM}
If we assume only one species of massive neutrinos, adiabatic
fluctuations give \cite{bi:bardeen}
\begin{equation}
T_{\nu , {\rm ad}}(k)={\rm exp} \left[ -0.16(kR_{f\nu})-(kR_{f\nu})^2/2 \right]
\left[1+1.6q+(4.0q)^{3/2}+(0.92q)^2\right]^{-1}\;,
\end{equation}
where
\begin{equation}
q\equiv \frac{k}{\Omega_{\nu} h_0^2{\rm Mpc}^{-1}}\qquad{\rm and}\qquad
R_{f\nu}=2.6(\Omega_{\nu}h_0^2)^{-1}{\rm Mpc}\;.
\end{equation}
We will take $\Omega_{\nu}=0.3$ and in the adiabatic CDM
transfer function we set $\Omega_{\rm CDM}=0.7$ .
The total power spectrum is then given by
$P(k)=\left(0.3\sqrt{P_{\nu}(k)}+0.7\sqrt{P_{\rm CDM}(k)}\right)^2$.\\ 
The results is $\delta_{\rm corr}(t_0)=-6.95 \cdot 10^{-4}$
 and the age of the universe becomes
\begin{equation}
t_0 \approx 0.99986 \cdot \frac{2}{3H_0}\;.
\end{equation}
\subsection{Warm Dark Matter Fluctuations}
Adiabatic fluctuations of warm dark matter give \cite{bi:bardeen}
\begin{equation}
T_{\rm warm, ad}(k)\approx 
\exp\left[-\frac{kR_{\rm fw}}{2}-\frac{(kR_{\rm fw})^2}{2}\right]
T_{\rm CDM, ad,}(k)\;,
\end{equation}
where
\begin{equation}
T_{\rm CDM, ad,}(k)=\left[1+1.7q+(4.3q)^{3/2}+q^2\right]^{-1}
\end{equation}
and
\begin{equation}
q\equiv \frac{k}{\Omega_{\rm CDM}h_0^2{\rm Mpc}^{-1}}
\qquad{\rm and}\qquad
R_{\rm fw}=0.2\left(\frac{g_{\rm CDM, dec}}{100}\right)^{-4/3}
(\Omega_{\rm CDM} h_0^2)^{-1}{\rm Mpc}\;.
\end{equation}
Here $g_{{\rm CDM, dec}}$ is the effective number of particle 
degrees of freedom when
the CDM particles decoupled, values range from 60-300, we will set 
$g_{{\rm CDM, dec}}=300$ and $\Omega_{\rm CDM}=1\,$.\\
The result is $\delta_{\rm corr}(t_0)=-2.90 \cdot 10^{-3}$ 
corresponding to an  age of the universe of
\begin{equation}
t_0 \approx 0.99942 \cdot \frac{2}{3H_0}\;. 
\end{equation}
\subsection{String and Texture Models}
The power spectrum of a cosmic string network evolving in a flat universe
dominated by CDM is given by \cite{bi:fisher,bi:albrecht}
\begin{equation}
P(k)=\frac{Ak}{\left[1+\alpha_2 k+(\alpha_3 k)^2+(\alpha_4 k)^3\right]
\left[1+(\alpha_5 k)^2\right]^2}\;,
\end{equation}
where $\alpha_2=7.57 h_0^{-2}$, $\alpha_3=5.89 h_0^{-2}$, 
$\alpha_4=1.93 h_0^{-2}$ and $\alpha_5=0.000357 h_0^{-2}$. \\
This power spectum assumes that the string network is characterized by
a scaling solution and that the power is dominated by the coherent
motions of loop strings; perturbations induced by string loops are
neglected \cite{bi:fisher}.\\
The power spectrum of a CDM universe with perturbations seeded by textures
is given by \cite{bi:turok,bi:peacock}
\begin{equation}
P(k)=\frac{Ak}
{\left[1+(\alpha k +(\beta k)^{3/2}+(\gamma k)^2)^{\nu}\right]^{2/\nu}}\;,
\end{equation}
with $\nu=1.2$, $\alpha=19.4 h_0^{-2}$, 
$\beta=6.6 h_0^{-2}$, and $\gamma=3.0 h_0^{-2}$,
where we still set $\Omega_0=1$.\\
Although the non-Gaussian nature of the string and texture models
means that the power spectrum does not provide a full description of
the density field even in the linear regime, the power spectrum is
sill a well-defined quantity, and its meaningful to compare it to
observations \cite{bi:fisher} and give excellent fits to it
\cite{bi:peacock}.\\
The result for the string and texture models are almost exactly the same
and give
$\delta_{\rm corr}(t_0)=-8.88 \cdot 10^{-3}$.
 The age of the universe in this case becomes
\begin{equation}
t_0 \approx 0.9982 \cdot \frac{2}{3H_0}\;.
\end{equation}
\section{Comparison with previous works}
\subsection{Newtonian treatment}
For a comparison with the Newtonian treatment by Buchert and Ehlers
\cite{bi:buchehlers,bi:ehlersbuch} we have to identify their
$\nabla{\bf v}$ with our $u^i_{\ ;i}$ and
their $\nabla{\bf u}$ with our $V^k_{\ k}$. Their result
\begin{equation}
\frac{\ddot a_D}{a_D}=-\frac{4\pi G}{3}\frac{M}{V}+
\frac{1}{3}<(u^k_{\ ,k})^2-u^k_{\ ,m}u^m_{\ ,k}>
\end{equation}
is found to agree exactly with ours in eq.~(\ref{eq:ddotaD}),
except for the derivatives, which are covariant in our case.
Eq.~(23) in \cite{bi:ehlersbuch} is an extension to describe
a globaly anisotropic universe.
They concluded, that their eq.~(9) in \cite{bi:buchehlers} and
eq.~(19) in \cite{bi:ehlersbuch} must also hold in general relativity,
because they were derived by averaging the Raychaudhuri equation.
In our treatment the Raychaudhuri equation can be derived by combining
eq.~(\ref{eq:G00}) and the trace of eq.~(\ref{eq:Rij}), replacing 
$u^i_{\ ;j}=(1/3)\Theta\delta^i_{\ j}+\sigma^i_{\ j}$.
So it is possible to recover all their results, exept for the
vorticity, which we assume to vanish in our treatment.
\subsection{Futamase's approximation scheme}
Futamase \cite{bi:futa,bi:futarev} calculated the backreactions based on his 
approximation scheme, where he introduced two small parameters
representing the amplitude of the metric fluctuations and the ratio
between the scale of the variation of this metric fluctuations and the
scale of a(t) and the background metric. Then in \cite{bi:futa}
he used a cosmological
post-Newtonian approximation. 
In \cite{bi:futarev} he employed the $3+1$ splitting of space time,
then  the Isaacson averaging \cite{bi:isaacson} is performed on the
background spatial hypersurface.
His results in his
eq.~(3.16) \cite{bi:futa} or in eq.~(68) \cite{bi:futarev} 
 are of the same order as
ours, but the factors are different.
There are several reasons for that discrepancy:
firstly in his approximation sheme he neglected some terms, we don't
use such an approximation. Secondly he did not introduce a scale
factor $a_D(t)$ defined by the expansion of a comoving volume. He
introduced the conformal factor $a(t)$, then he rescaled this scale
factor by neglecting terms like $<{\bar h}^k_{\ k}>$. Thirdly his 
averaging process in \cite{bi:futa} is not defined using the
square-root of the real metric under the integral, rather he used the
square-root of the background metric, which is essentially unity for a
flat background.
He also used this averaging process at the end of \cite{bi:futarev}
to recover the results of \cite{bi:bild}.
\subsection{Bildhauer and Futamase}
Bildhauer and Futamase \cite{bi:bild} calculated the backreactions of
the inhomogeneities based on the work of
Futamase \cite{bi:futa} and the Newtonian Zel'dovich approximation
(Buchert \cite{bi:buch}).
Their result (eq.~(25), see also eq.~(84) in \cite{bi:futarev}) reads
with $b\equiv\delta_{\rm corr}(t_0)$
\begin{equation}
\delta_{\rm corr}(t_0)
=\frac{19}{36}10^{-6}h_0^2(1+z_{\rm in})^4\frac{1}{\mu^2}<|\vec U|^2>\;,
\end{equation}
where we want to indicate a typing error ($M_1$ defined in their eq. (25) is
not the same as in their eq. (22), the factor $k^2/\mu^2$ is 
incorporated into $M_1$).
With $\mu^{-2}<|\vec U|^2>= \mu^{-2}<|\nabla s_{\rm in}|^2> =
<\Psi^{,m}\Psi_{,m}>$ this is of the same order as our result,
only the factor is different.
The reasons are the same as those in the previous subsection.
Another error was found by Futamase \cite{bi:futarev}: 
 $M_1=57\pi^3$ should read $M_1=57/8$, the mistake is coming from
the wrong integration region $[0,2\pi]$ instead of $[0,d]$. 
They derived at the conclusion that the underestimation of the age of
the universe is approximately 30 percent, which is not correct since
they just assumed $\delta_{\rm corr}(t_0)$ to be of order unity instead of
calculating it quantitatively as we did here.
\section{Concluding remarks}
We have calculated quantitatively the influence of the inhomogeneities
on the global expansion factor of a flat universe with vanishing
cosmological constant in the framework of a Zel'dovich type relativistic
approximation scheme using the results from COBE.
The first result is that the backreactions act as an additional energy
density, which is proportional to $a_D^{-2}$, so we can interpret the averaged
expansion as Friedmannian with a small positive spatial curvature.
The second result is that this influence is very small.
As a consequence of this the modification of the age
of the universe calculated in the usual way 
(i.e. assuming a homogeneous universe) with a given Hubble constant
is negligible. In all models considered here relative differences were
less than $\approx 2 \cdot 10^{-3}$.
This does not imply that the inhomogeneities are negligible for local
astronomical measurements of the Hubble constant.
Locally the determination of the redshift-distance relation
can be strongly influenced by the peculiar velocity fields
due to inhomogeneities ($d=H z+{\cal O}(2)$).
Our calculation does not consider
such effects, but is contrained to comparing globally homoeneous
and averaged inhomogeneous matter distributions.
Calculating the modification of the redshift-distance relation will
be the subject of future investigations.
As a result the age problem of the universe that arises in high
density models can only be solved either
with a lower Hubble constant, with a nonzero cosmological constant
or with a reduced age of globular clusters.
\acknowledgments

H. R. wants to thank H. Riffert and A. Geyer for stimulating and helpful
discussions.
\newline
The authors would like to thank T. Buchert and J. Ehlers for valuable
comments.
\newline
H. R. was supported by the ``Deutsche Forschungsgemeinschaft".
\appendix
\section{Commutation rule}\label{A}
The time derivative of an averaged quantity reads:
\begin{equation}
\frac{d}{dt}<A>=
-\frac{\dot V}{V}<A>
+\frac{1}{V}\int_{V} ({\dot A}\sqrt{g}+A{\dot{\sqrt{g}}})d^3x\;.
\end{equation}
This leads to the commutation rule 
\cite{bi:explain,bi:buchehlers,bi:ehlersbuch}
\begin{equation}
\frac{d}{dt}<A>-<\dot A>
=-<\theta><A>+<A\theta>\;,
\end{equation}
where
\begin{equation}
\theta=\frac{\dot{\sqrt{g}}}{\sqrt{g}}
\qquad {\rm and} \qquad
<\theta>=3\frac{\dot a_D(t)}{a_D(t)}\;.
\end{equation}
To convert eqs.~(\ref{eq:dota}) and (\ref{eq:ddota})
to the eqs.~(\ref{eq:dotaD}) and (\ref{eq:ddotaD}) we used
\begin{equation}
\frac{d}{dt}<V^k_{\ k}> - <\dot V^k_{\ k}> =
<(V^k_{\ k})^2> - <V^k_{\ k}>^2
\end{equation}
and neglected the term $<V^k_{\ k}>^2$, because it is
a higher order quantity.
\section{Model of the inhomogeneous universe}\label{B}
The result of the relativistic Zel'dovich approximation to second
order \cite{bi:russ} is the following metric tensor:
 
\begin{eqnarray}\label{solution}
\gamma_{ij} &=&   
  \left(1 + \frac{20}{9\,c^2t_{\rm in}^2}\Psi \right)
    \delta_{ij} + \,2a_D(t)\Psi_{,ij}\nonumber\\
 && + \,\frac{a_D(t)}{c^2t_{\rm in}^2}\left[
                -\frac{20}{3}\Psi_{,i}\Psi_{,j}
                -\frac{40}{9}\Psi\Psi_{,ij} 
                +\frac{10}{9}\Psi^{,k}\Psi_{,k}\,\delta_{ij} 
          \right] \\
 && + \,a^2_D(t)\left[ \frac{19}{7} \Psi^{,k}_{\ ,i}\Psi_{,kj} 
                  -\frac{12}{7} \Psi^{,k}_{\ ,k}\Psi_{,ij} 
                  +\frac{3}{7}\left(\left(\Psi^{,k}_{\ ,k}\right)^2
                    - \Psi^{,k}_{\ ,\ell}\Psi^{,\ell}_{\ ,k} \right)
                   \delta_{ij} 
            \right]  \;, \nonumber
\end{eqnarray}
where we set $\alpha=-50/81$. Since their difference
is only of second order, we could replace $a(t)$ by $a_D(t)$.
The determinant we only need to first order:
\begin{equation}
\sqrt{\gamma}=1
+\frac{10}{3\,c^2t_{\rm in}^2}\Psi
+a_D(t)\Psi^{,k}_{\ ,k}\;.
\end{equation}
This leads to
\begin{eqnarray}
<{^3\!R}>&=&\frac{1}{a_D^2c^2t^2_{\rm in}V_{\rm in}}\int_{V}
\Biggl(-\frac{40}{9}\Psi^{,m}_{\ ,m}
+a_D(t)\frac{20}{9}\Bigl(\Psi^{,k}_{\ ,m}\Psi^{,m}_{\ ,k}
-\left(\Psi^{,k}_{\ ,k}\right)^2\Bigr)\nonumber\\
&&+\frac{1}{c^2t_{\rm in}^2}\left(\frac{400}{81}\Psi\Psi^{,m}_{\ ,m}
+\frac{600}{81}\Psi^{,k}\Psi_{,k}\right)\Biggr)d^3x
\end{eqnarray} 
and
\begin{equation}
<(V^k_{\ k})^2-V^k_{\ m}V^m_{\ k}>
=\frac{4}{9a_Dt^2_{\rm in}V_{\rm in}}
\int_{V}\left((\Psi^{,m}_{\ ,m})^2
-\Psi^{,k}_{\ ,m}\Psi^{,m}_{\ ,k}\right)d^3x\;.
\end{equation}
The averaged Friedmann equations read:
\begin{eqnarray}
\frac{\dot a_D^2}{a_D^2}&=&\frac{8\pi G}{3c^2}<\rho>+
\frac{1}{t^2_{\rm in}V_{\rm in}}\int \Biggl(
\frac{20}{27a_D^2}\Psi^{,m}_{\ ,m}
+\frac{8}{27a_D}\left((\Psi^{,m}_{\ ,m})^2
-\Psi^{,k}_{\ ,m}\Psi^{,m}_{\ ,k}\right)\nonumber\\
&& -\frac{1}{a_D^2c^2t^2_{\rm in}}
\left(\frac{300}{243}\Psi^{,m}\Psi_{,m}
+\frac{200}{243}\Psi\Psi^{,m}_{\ ,m}\right)\Biggr)d^3x
\end{eqnarray}
and
\begin{equation}
\frac{\ddot a_D}{a_D}=-\frac{4\pi G}{3c^2}<\rho>
+\frac{1}{t^2_{\rm in}V_{\rm in}}\int
\frac{4}{27a_D}\left((\Psi^{,m}_{\ ,m})^2
-\Psi^{,k}_{\ ,m}\Psi^{,m}_{\ ,k}\right)d^3x\;.
\end{equation}
Integration and the assumption of periodic boundary conditions lead to
\begin{equation}
\int_V \Psi^{,m}_{\ ,m}d^3x=0\;.
\end{equation}
The Fourier transformation
$\delta({\bf x},t_{\rm in})
=\sum_{\bf k}\delta_{\bf k}e^{i{\bf kx}}$
together with 
$\int_V {\rm exp}^{i({\bf k-k'}){\bf x}}d^3x
=\delta_{\bf kk'}$
leads to
\begin{equation}
\int_{V}\left(\Psi^{,m}_{\ ,m}\right)^2d^3x
=\int_{V}\Psi^{,m}_{\ ,k}\Psi^{,k}_{\ ,m}d^3x
\end{equation}
and
\begin{equation}
\int_{V}\Psi\Psi^{,m}_{\ ,m}d^3x=-\int_{V}\Psi^{,m}\Psi_{,m}d^3x\;.
\end{equation}
The averaged density is treated as follows:
\begin{equation}
\frac{8\pi G}{3c^2}<\rho>
=\frac{8\pi G}{3c^2}\frac{1}{V}
\int\rho(t_{\rm in})\sqrt{g(t_{\rm in})}d^3x
=\frac{8\pi G}{3}\frac{M}{V}\equiv\frac{8\pi G}{3c^2}\rho_b(t)\;,
\end{equation}
where we used the conservation law 
$\rho(t_{\rm in})\sqrt{g(t_{\rm in})}=\rho(t)\sqrt{g(t)}$.
Note that even in a case where the averaged second scalar invariant
would not vanish our treatment would still be consistent, since in every case
\begin{equation}
\frac{d}{dt}\left(\frac{\dot a_D^2}{a_D^2}\right)=
2\frac{\dot a_D}{a_D}
\left(\frac{\ddot a_D}{a_D}-\frac{\dot a_D^2}{a_D^2}\right)
\end{equation}
is satisfied.

\end{document}